\documentclass[aps,prmaterials,notitlepage,superscriptaddress,reprint]{revtex4-2}
\usepackage[final]{graphicx}
\usepackage{natbib}
\usepackage{hyperref}
\usepackage{amssymb}
\usepackage{amsmath}
\usepackage{mathrsfs}
\usepackage{xcolor}
\usepackage{bm}
\begin{document}
\title{
Computational tuning of the elastic properties of low- and high-entropy ultra-high temperature ceramics
}
\author{Samuel J. Magorrian}
\affiliation{Hartree Centre,  STFC Daresbury Laboratory, Warrington WA4 4AD, United Kingdom}
\author{Ljiljana Stojanovi\'c} 
\affiliation{Hartree Centre, STFC Daresbury Laboratory, Warrington WA4 4AD, United Kingdom}
\author{Lara Kabalan} 
\affiliation{Hartree Centre,  STFC Daresbury Laboratory, Warrington WA4 4AD, United Kingdom}
\author{Ardita Shkurti}
\affiliation{Hartree Centre,  STFC Daresbury Laboratory, Warrington WA4 4AD, United Kingdom}
\author{Richard N. White}
\affiliation{Lucideon Limited, Stone Business Park, Brooms Road, Stone, Staffordshire, ST15 0SH, UK}
\author{Fabian L. Thiemann} 
\affiliation{IBM Research Europe,
	Daresbury, WA4 4AD, UK}
\author{Viktor Z\'olyomi}
\affiliation{Hartree Centre, STFC Daresbury Laboratory, Warrington WA4 4AD, United Kingdom}

\begin{abstract}
Ultra-high temperature ceramics (UHTCs) represent a class of crystalline materials for extreme environments. They can withstand extremely high temperatures but are mechanically difficult to work with due to their inherent brittleness. Mixture compounds, in particular high-entropy mixtures, offer a pathway to tune the physical properties of UHTCs such as their elastic constants. Here we fine-tune the MACE-MPA-0 universal machine-learning potential on rocksalt carbide UHTCs containing group IV–V metals and demonstrate that not only do the elastic constants deviate from the rule of mixtures approximation in the high-entropy limit, but also in the low-entropy limit of binary and ternary mixtures. We find that this is caused by distortion imposed by the lattice mismatch, enabling the tuning of the physical properties of UHTC mixtures in both low- and high-entropy compounds. We identify a three-component mixture compound, HfCVCZrC, as the best balance between synthesizability and toughness, and apply our developed MACE-UHTC model to identify a range of non-equimolar candidate compositions of this compound which may enable the synthesis of a mixture UHTC with a Young's modulus up to 40 GPa below that of ZrC.
\end{abstract}

\maketitle

\section{Introduction}

Entropy stabilization of crystals has emerged in recent years as a way to engineer new material compositions \cite{Zhang2019_HEC_Review_C9TA05698J,AKRAMI2021100644,YANG2025177691,Liu2025_HEC_Review}. The core idea is that by mixing together 4 or more elements in a crystal, the configuration entropy rises above a threshold where it is defined as a high-entropy crystal. Beyond enabling the coexistence of otherwise unstable element combinations through high configuration entropy, this unlocks the exciting opportunity to precisely tune and tailor material properties for specific applications. For example, high-entropy thermal barrier coatings \cite{ZHANG20221349} can be developed with desired thermal expansion coefficients to match those of protected materials, while also substantially reducing the thermal conductivity compared to the rule of mixtures average due to increased phonon scattering caused by the high-entropy mixing.

A particular class of materials that benefit greatly from high-entropy mixture design are ultra-high temperature ceramics (UHTCs) \cite{Castle2018}. These are a unique class of ceramics that can withstand extremely high temperatures in reactive environments. This makes them attractive candidates for applications in the energy and aerospace sectors, for instance as structural materials in fusion reactors, as well as coatings for the exterior of hypersonic craft. The low deformability of these materials, 
 however, reduces their resilience to wear and tear.

A possible route to improved mechanical performance of UHTCs is a reduction in the Young's modulus. This improves the crystal's ability to absorb and redistribute stress, enhancing toughness. For systems that are subject to rapid heating such as may be anticipated for high-entropy ceramic coatings this is definitely beneficial. The Young's modulus of technical ceramics typically falls in the 200 to 450 GPa range \cite{NIST_ElasticModData,Book_CeramicMechanics}. For the single metal UHTCs we study here we find Young's moduli in the range 400 to 500~GPa, making them more sensitive to thermal shock and resulting in reduced toughness. A reduction by even a few 10s of GPa in a mixture compound could equate to a significant improvement in toughness.
 
High-entropy compositions could overcome the limitations of single-component UHTCs, as high configurational entropy may stabilize mixtures that would otherwise not be synthesizable. They can also result in sufficient configurational disorder to substantially break the rule of mixtures, which enables significant tuning of physical properties by changing the composition.
Significant improvement of yield and failure strength has been demonstrated in a four-component high-entropy compound\cite{Csanadi2019} without an increase in brittleness.
The enhanced plasticity in high-entropy refractory ceramics was explained by the increasing valence electron concentration in the high-entropy compound\cite{SANGIOVANNI2021109932}. Computational predictions showed that high-entropy compositions increase hardness due to dislocation slips being less likely when atomic randomness increases at the dislocation core, confirmed in nanoindentation experiments where hardness improved by 25\% compared to the rule of mixtures in eight-component ceramics\cite{Wang2020_HEC_Peierls_DFT}. The critical question now is how to identify the best compositions of high-entropy UHTCs for improved mechanical performance, within a vast composition space.

While fundamental work has established the synthesizability of equimolar high-entropy ceramics through density functional theory (DFT) simulations\cite{Divilov2024} relying on the partial occupancies (POCC) method \cite{Yang2016POCC}, their elastic properties remain largely unexplored at first-principles accuracy due to the computational cost of computing the elastic tensor for configurationally disordered materials where ensemble averaging over many configurations is necessary. The cost of predicting thermal expansion is even greater. It is possible to avoid ensemble averaging through special quasirandom structures (SQSs) \cite{Zunger1990SQS,WANG2021128754} but large SQSs are needed for accurate predictions in high-entropy compounds due to the large number of component elements in the crystal \cite{Gao2016}. In addition, true fine-tuning of the properties of high-entropy compounds requires considering non-equimolar mixtures where DFT would especially struggle, even with the SQS or POCC methods, as non-equimolar compositions increase the required size of the simulated structure.

An alternative approach is to use the coherent potential approximation (CPA) \cite{CPA_PhysRevB.5.2382} or the virtual crystal approximation (VCA) \cite{VCA_PhysRevB.61.7877}. Within these approximations a mean field pseudoatom potential is used to describe the random mixture compound. This enables representation of the disordered crystal just by the primitive cell alone. Once the CPA or VCA potential for a mixture is constructed, calculations of elastic or thermomechanical properties are affordable, even for non-equimolar compounds. A key approximation relied on is that all local distortions can be neglected. In metal alloys where the atomic radii are comparable\cite{Huang_2022,MingLong2025}, this approximation is sound, but when component elements differ in size or chemistry significantly \cite{WANG2021128754} or in high-entropy UHTCs or other ceramics with multiple sublattices, lattice distortion can be substantial. A method which accounts for local distortion without incurring the computational expense of DFT calculations on disordered supercells would be preferable.

Recent advances in machine learning interatomic potentials (MLIPs) now offer a practical solution to the high computational cost of first-principles calculations of elastic properties \cite{Zuo-2020, Unke-2021,behler2007, Behler-2016, Thiemann-2024, Jacobs-2025, Friederich-2021, Deringer-2019}. By learning from large databases of atomistic configurations labeled with energies, atomic forces, and stress tensors obtained from \textit{ab-initio} calculations, MLIPs can accurately reproduce the underlying physics at a fraction of the computational expense. This enables the efficient and systematic evaluation of elastic properties in high-entropy UHTCs with near-first principles accuracy, enabling large-scale studies that were previously infeasible.

Equivariant message-passing networks have become the state-of-the-art models for atomistic simulations, as they inherently respect the symmetries of physical systems. Among these, the MACE architecture \cite{Batatia2022mace} has emerged as a leading approach, combining high accuracy and efficiency with strong interpolation capabilities in low-data regimes by incorporating higher-order body messages. This has recently culminated in robust pretrained models like MACE-MP-0, trained on the MPtrj dataset from the Materials Project \cite{Horton2025}, and MACE-MPA-0, which additionally used the sAlex dataset (subsampled dataset derived from the Alexandria database\cite{AlexandriaDataset}), thus encompassing a total training dataset of over 1.6 million configurations of inorganic crystals computed with DFT. While these models enable stable molecular dynamics and qualitative agreement with \textit{ab-initio} methods, and can even be used for synthesizability predictions of high-entropy oxide crystals \cite{Dicks_HEO_2025}, fine-tuning to the system and level of theory of interest remains essential to gain mechanistic and quantitative insight at first principles accuracy\cite{Elena2025}.

In this work we explore, using a model that approaches first principles accuracy, how the elastic properties of multi-component UHTCs vary with composition.
We finetune MACE-MPA-0 for UHTC crystals spanning an extensive set of over 60 equimolar mixtures—ranging from simple binaries to complex high-entropy systems. We perform ensemble-averaged calculations of the mechanical properties of all possible equimolar mixtures, and apply the model to non-equimolar mixture compounds. We observe substantial deviations from the rule of mixtures, even in the binary case. We attribute this to significant lattice distortions, which correlates strongly with the lattice parameter mismatch between the components in the mixture compounds, leading to the breakdown of the rule of mixtures irrespective of the magnitude of the configurational entropy. 
This suggests the exciting potential to tune the elastic properties of UHTCs, even in the low-entropy limit.

We use our model to scour the composition space for compounds that exhibit low Young's moduli and reasonable anticipated synthesizability. We identify a three-component mixture, HfCVCZrC, as the most promising candidate among the equimolar mixtures. This has a Young's modulus 10~GPa less than the smallest available among the rocksalt carbide components (that of ZrC). We show that further optimization is possible by tuning the composition away from the equimolar limit, reducing its Young's modulus by up to an additional 30~GPa.

\section{Results}
\begin{figure*}
    \centering
    \includegraphics[width=1.0\textwidth]{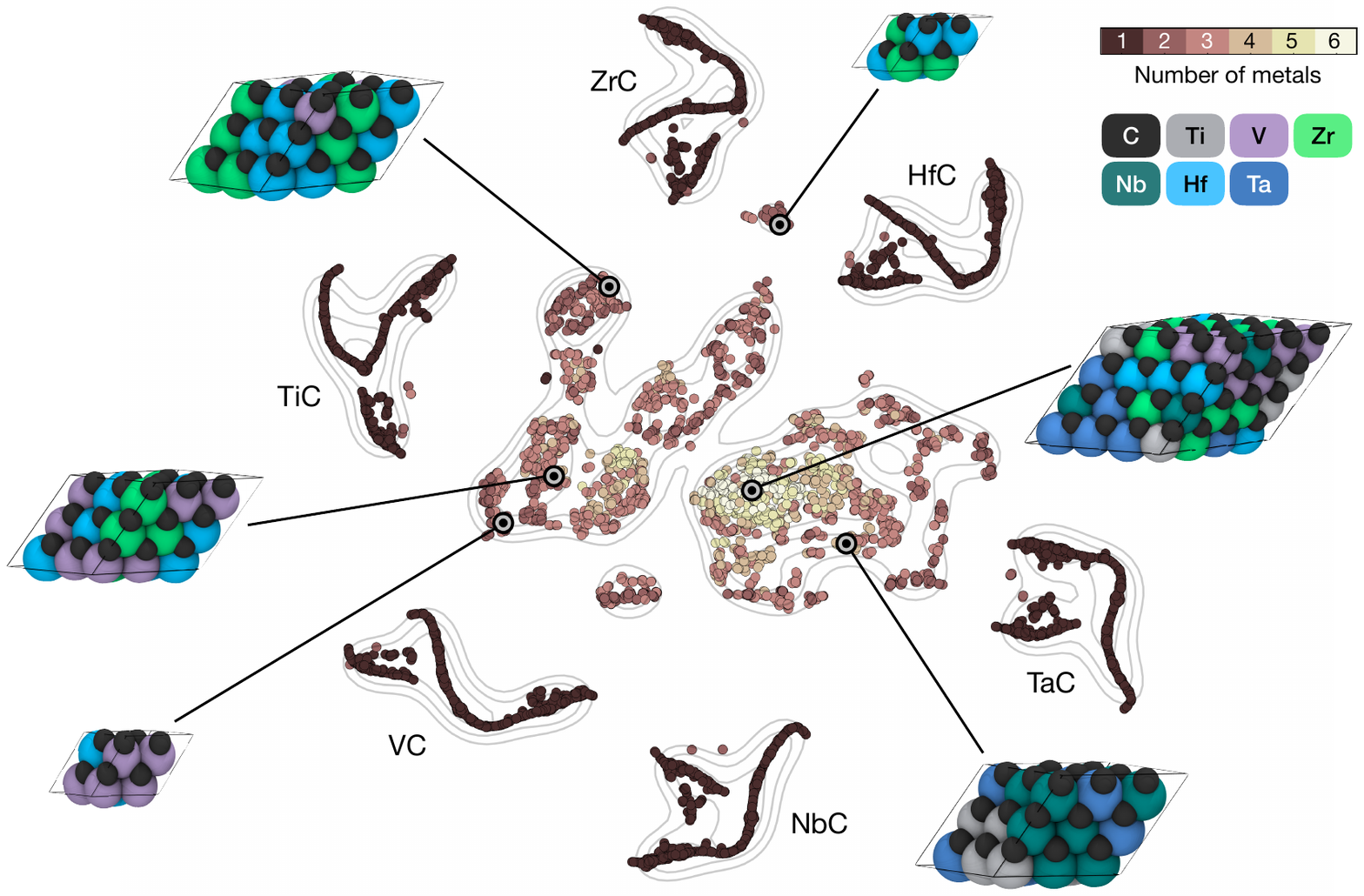}
    \caption{Overview of the chemical space covered by the finetuned MACE-UHTC model, shown through t-SNE projections of configuration-averaged MACE embeddings of the training data. Each point represents an atomic configuration and is colored by the number of metal species present, with example UHTC crystal structures displayed in the insets. Contour lines indicate regions of equal density.
    }
    \label{fig:umap_representation}
\end{figure*}
\subsection{The material space of UHTCs under investigation}

UHTCs are typically carbides, nitrides, carbonitrides, or borides of early transition metals. Here we focus on the subset of rocksalt carbides, which readily exist with either group IV or group V metals on the cationic sublattice. This defines a realistic chemical space for which to finetune a MACE model \cite{Batatia2022mace} comprising 6 transition metals (Ti, Zr, Hf, V, Nb, Ta) in addition to carbon on the anionic sublattice. The list of possible equimolar mixtures includes 15 possible two-component, 20 three-component, 15 four-component, 6 five-component, and 1 six-component mixtures. Along with 6 single-component materials, this results in a total of 63 distinct equimolar material compositions. We systematically investigate all of these using MACE-UHTC, the MLIP fine-tuned in this work. As we will show, HfCVCZrC emerges as the most promising mixture compound, which we therefore subject to a full analysis of its non-equimolar composition space.

To further characterize this chemical space and to understand how the model represents structural diversity, we analyzed the learned MACE atomic environments across all configuration types. Atomic embeddings were extracted from the second interaction layer of MACE-UHTC, averaged over atoms within each configuration of the training set, and projected into two dimensions using t-SNE\cite{JMLR:v9:vandermaaten08a} (Fig.~\ref{fig:umap_representation}). Single-metal carbides form distinct, well-separated groups, while mixed-metal carbides (2TM–6TM) occupy the intermediate regions. A gradual shift of the embeddings is observed, from carbides with two metallic centers to higher-order multi-component systems, indicating that the model systematically captures increasing structural and chemical complexity. The 2TM carbides interpolate between pairs of single-metal carbides and thus span over larger values of the projected t-SNE dimensions, whereas 3TM-6TM systems become progressively more centered as they interpolate among multiple components.

\begin{figure*}
    \centering
    \includegraphics[width=1.0\textwidth]{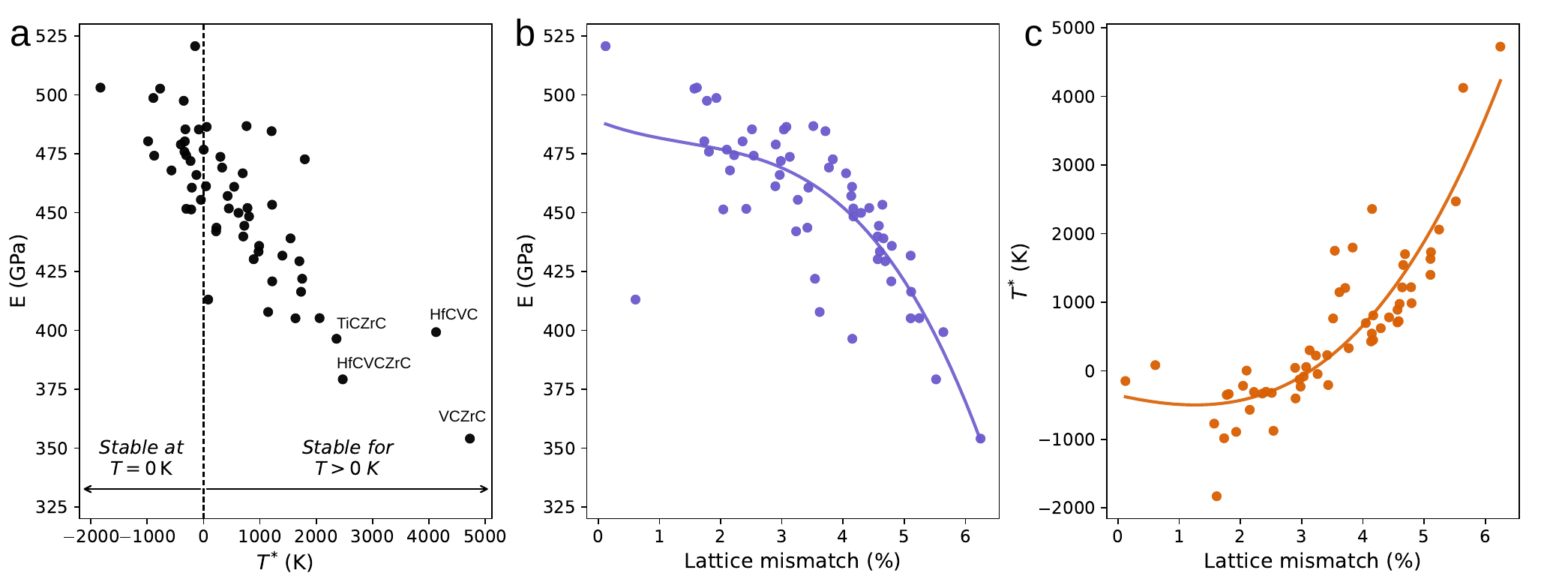}
    \caption{High-throughput screening of equimolar compounds for Young's modulus and synthesizability. (a) Pareto plot showing Young's modulus ($E$ in GPa) versus the effective stabilization temperature ($T^{*}$ in K). (b) Dependence of Young's modulus on lattice mismatch. (c) Dependence of effective stabilization temperature on lattice mismatch. Curves in (b) and (c) are guides to the eye.}
    \label{fig:666_moduli_vs_mismatch}
\end{figure*}

\subsection{Synthesizability estimation through the effective stabilization temperature}

High-entropy compounds are entropy-stabilized mixture compounds where solid solutions of multiple crystals are stabilized by the configurational disorder arising from the random distribution of the component elements across the lattice. The mixing entropy increases with the number of possible components in the mixture compound. This is important as the mixing entropy multiplied by the negative of the temperature adds a contribution to the free energy. As the crystal suffers an energy penalty from mixing together components that would prefer to segregate, this penalty can be countered by the mixing entropy as the temperature increases. This allows a very simple approximation by which synthesizability can be ranked in these materials by identifying the temperature where the mixing entropy contribution fully balances out the energy penalty from the mixing. This quantity we call the effective stabilization temperature, $T^{*}$. This is not a true temperature but an effective temperature, as it takes negative values for mixture compounds that are already energetically favorable even without the mixing entropy contribution, i.e. when the mixing energy is negative. It is a quantity that is easy to obtain from MACE-UHTC calculations and can indicate how difficult it might be to synthesize each mixture compound.

In Fig.~\ref{fig:666_moduli_vs_mismatch}a we show a Pareto plot of $T^{*}$ versus the predicted Young's modulus for all equimolar compounds in the investigated composition space. It is immediately apparent that some mixtures are already stable without the configurational entropy contributions, while some suffer from substantial energy penalties which result in a very large $T^{*}$ value. The two compounds with the largest $T^{*}$ are HfCVC and VCZrC, which aligns with existing predictions that these two mixtures cannot exist as single-phase crystalline solids according to the order-parameter functional model \cite{Gusev_2000}. The situation improves substantially when these two are combined into an equimolar mixture of HfC, VC, and ZrC: the increased mixing entropy and, to a smaller extent, the decrease in mixing energy (due to a smaller lattice mismatch in the ternary system compared to VCZrC and HfCVC, Fig. \ref{fig:666_moduli_vs_mismatch}) result in a decrease of $T^{*}$ by almost a factor of 2 in HfCVCZrC compared to the least stable mixture, VCZrC. The predicted $T^{*}$ for equimolar HfCVCZrC is just slightly larger than for TiCZrC, a binary carbide which has been successfully synthesized \cite{TiCZrC_2022}, indicating that the ternary carbide HfCVCZrC should be possible to make as well based on the predicted $T^{*}$ values.

The predicted lack of stability in HfCVC and VCZrC can be traced back to the large lattice mismatch between their respective components, which can be quantified by the ratio of the standard deviation ($\Delta a$) and the mean ($\bar{a}$) of the lattice parameters of the component carbides. In fact, as we show in Fig.~\ref{fig:666_moduli_vs_mismatch}b and Fig.~\ref{fig:666_moduli_vs_mismatch}c which plot the Young's modulus and $T^{*}$ against the lattice mismatch between the components, lattice mismatch drives both the reduction of the Young's modulus and the synthesizability in equimolar mixture UHTCs.

\section{Composition optimization with MACE-UHTC}
\begin{figure*}
    \centering
    \includegraphics[width=1.0\textwidth]{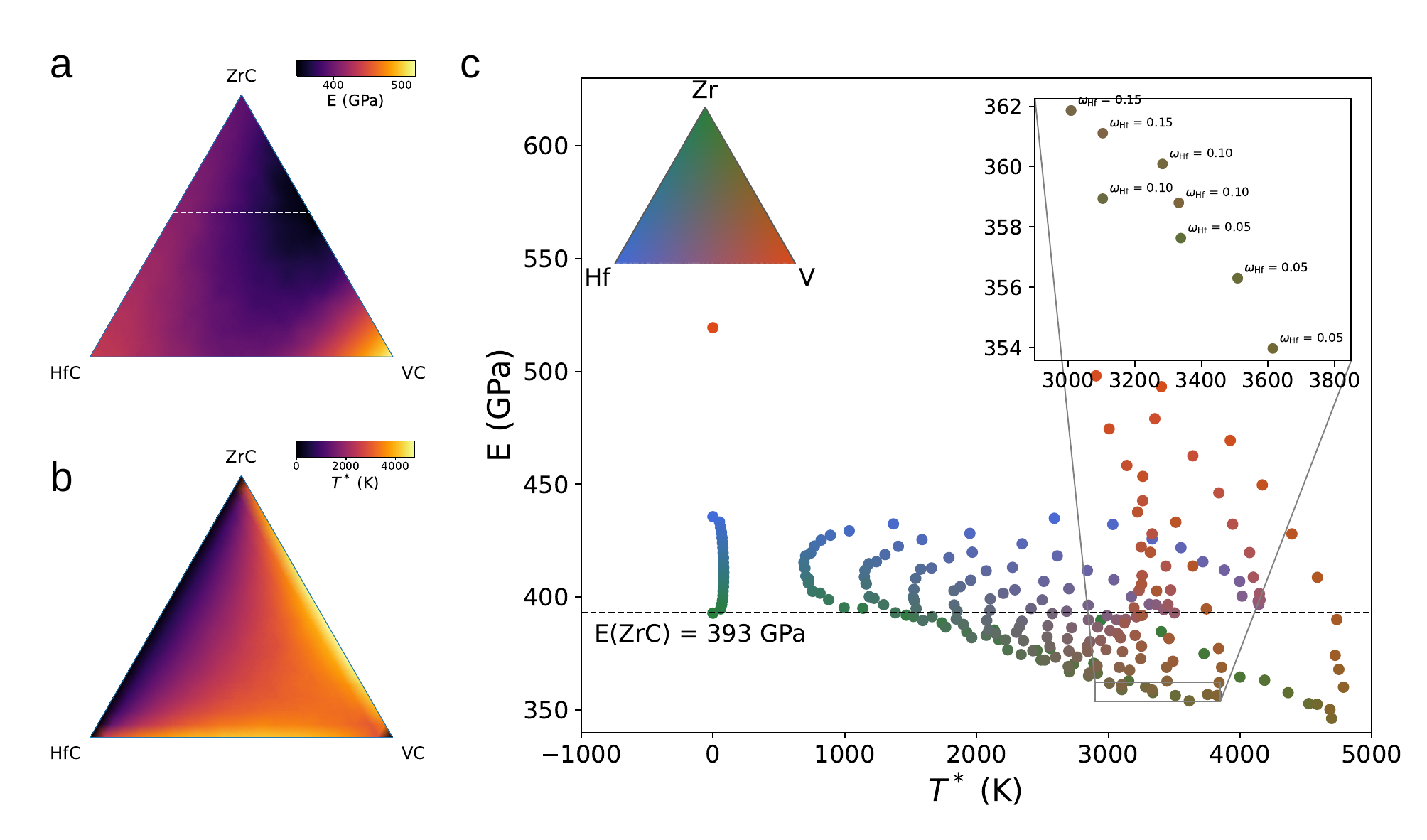}
    \caption{(a) Ternary diagram showing the dependence of the Young's modulus ($E$ in GPa) on the Hf-V-Zr composition in the HfCVCZrC compound. Dashed line indicates path in composition space plotted in Fig.~\ref{fig:discussions}(b). (b) Ternary diagram showing the dependence of the effective stabilization temperature ($T^{*}$ in K) on the Hf-V-Zr composition in the HfCVCZrC compound. (c) Predictions for the Young's modulus and the effective stabilization temperature in the full non-equimolar composition space for the HfCVCZrC mixture UHTC compound.}
    \label{fig:Pareto_HfVZr}
\end{figure*}
As mentioned in the Introduction, there is a strong need for mixture UHTCs with reduced Young's moduli. Having ruled out the unstable HfCVC and VCZrC mixtures, our search for equimolar mixtures the Young's modulus of which is smaller than the predicted 393 GPa of ZrC leads us to a single candidate compound, HfCVCZrC, with a Young's modulus of 380 GPa (Fig. \ref{fig:666_moduli_vs_mismatch}). A reduction of 13 GPa is very useful, but the question arises: can we optimize this compound further by adjusting its composition?

This is where the strength of the MACE-UHTC model really shows itself, as the model enables us to compute the elastic properties, as well as the effective stabilization temperature, in the full non-equimolar composition space of HfCVCZrC with a dense coverage. Fig.~\ref{fig:Pareto_HfVZr} shows the predicted Young's modulus and $T^{*}$ values in this continuum composition space. The Pareto front identifies multiple potential candidate compounds which exhibit a markedly greater reduction of the Young's modulus. The drawback is that the greater this reduction, the larger the $T^{*}$, which is due to the non-equimolar mixture slowly inching closer to the limit of a two-component mixture. We can reduce the Young’s modulus all the way down to 354 GPa but by then $T^{*}$ increases to 3600 K (Fig. \ref{fig:Pareto_HfVZr} inset). In these particular mixtures, the Hf content is reduced to 5\% which may or may not be sufficient to stabilize the ternary mixture, depending on the balance between V and Zr content.

However, along the Pareto front, our predictions provide a broad range of non-equimolar HfCVCZrC compositions for consideration. If the equimolar compound is eventually synthesized and its Young's modulus found to be in line with our prediction, the predicted Pareto front provides the guideline for how to get the most out of this mixture compound by optimizing its composition, as long as it remains stable as a single phase crystal.

\section{Discussion}

A characteristic feature of mixture compound UHTCs is that due to the configurational disorder, non-zero forces act on the atoms at every lattice site. This results in displacement of atoms from their ideal positions. The cell deformation compared to the rule of mixtures estimate, largely manifested as isotropic strain, is quite small ($< 0.7$\%), whereas the ionic displacement of atoms can exceed 0.1~\AA. Both the origin and the consequences of this lattice distortion warrant discussion.

\begin{figure}
    \centering
    \includegraphics[width=\linewidth]{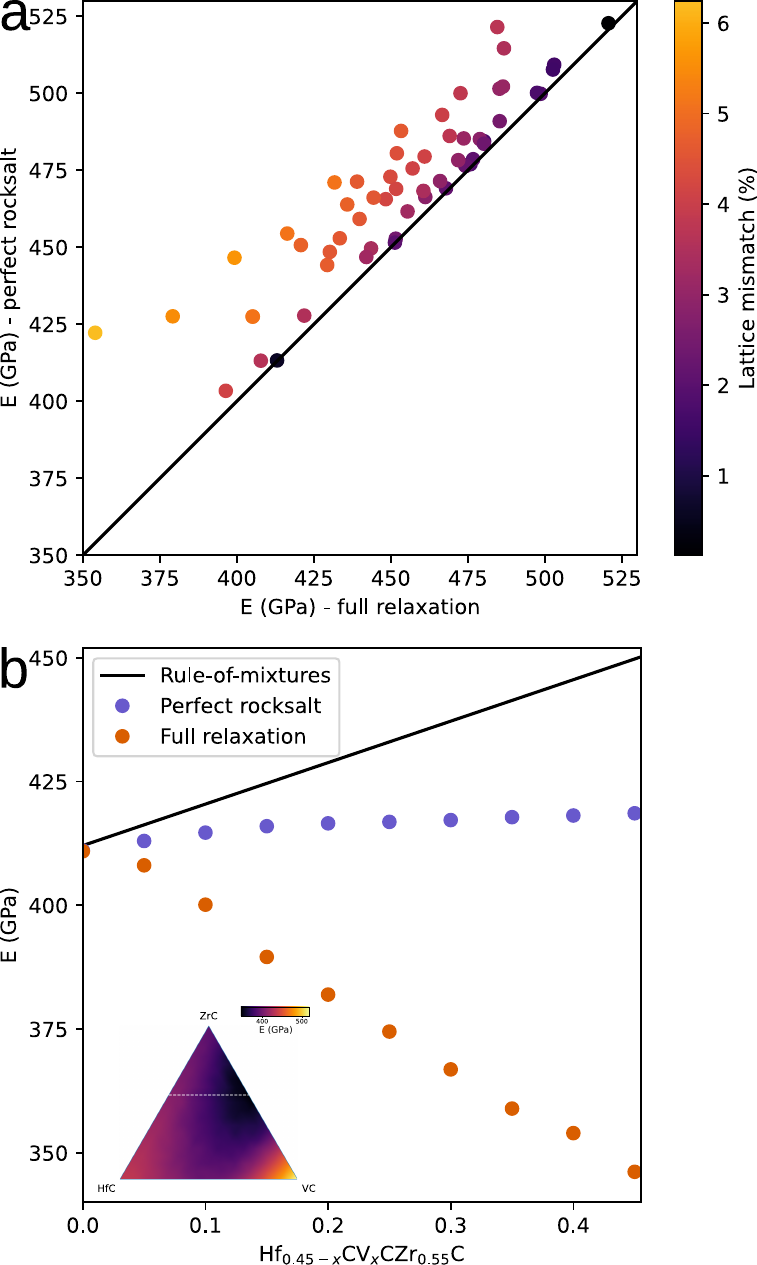}
    \caption{Demonstration of the importance of taking lattice distortions into account on (a) the Young's modulus of all equimolar compounds comparing the fully relaxed ensemble MACE-UHTC predictions to the (VCA-esque) approximation of a perfect rocksalt crystal in which only the volume is optimized (points colored according to the mismatch between the lattice constants of the component materials), and (b) the Young's modulus of HfCVCZrC along the path indicated in the inset according to the rule-of-mixtures and using MACE-UHTC with and without full relaxation.}
    \label{fig:discussions}
\end{figure}

We repeated the elastic constant calculations for all equimolar mixtures, and a selection of non-equimolar compounds along the Pareto front of HfCVCZrC (identified in Fig.~\ref{fig:Pareto_HfVZr}), by neglecting all ionic distortion and enforcing hydrostatic strain on the cell, which mimics methods such as the CPA or VCA. Fig. \ref{fig:discussions} clearly shows that the Young's modulus is overestimated when computed with this approximation. This further evidences the importance of taking ionic distortions into account in high-entropy UHTCs.

The clear significance of the lattice distortion demonstrates why models like MACE-UHTC are desperately needed in materials science today. Traditionally, VCA or CPA approaches were used to model non-equimolar mixture compounds at first principles accuracy, but for materials like UHTCs this approach is clearly not accurate enough. The alternative of using DFT with SQSs or the POCC method meanwhile limits applicability when it comes to non-equimolar mixtures due to the need for very large SQSs or POCC tiles. A MLIP fine-tuned for UHTCs overcomes both of these limitations, as the speed and scalability of MACE-UHTC combined with the first principles precision it exhibits enables the efficient modelling of any composition by simple ensemble averaging of many configurations in large supercells. 

Our approach to discovery of optimized mixture compound UHTCs through MACE-UHTC is powerful and efficient, but not without limitations. MACE-UHTC is trained on semi-local DFT data, and while DFT performs well for elastic constant prediction, it is not an exact method, hence errors compared to reality can be expected. Furthermore, we compute zero temperature elastic constants, which is an accurate approximation for ambient conditions, but the neglected impact of high temperature on the elastic constants limits our accuracy for making predictions at the intended temperature of operation for UHTCs. Temperature dependence could be accounted for by molecular dynamics simulations, which would incur a significant expense for high-throughput screening. Alternatively, one could compute temperature dependent elastic constants from the thermodynamic free energy in the quasiharmonic approximation, but MACE-UHTC can only compute the lattice contributions to the free energy, while the temperature-dependent electronic contributions would either have to be neglected or computed by DFT. It is also worth emphasizing that $T^{*}$, the effective stabilization temperature, while a simple measure that correlates well with much more expensive synthesizability descriptors (as discussed in the SI, section I), it is a crude approximation and does not correspond to a real temperature. The advantage of $T^{*}$ is that it is very easy to compute using MACE-UHTC and can therefore be used for rapid high-throughput screening, but $T^{*}$ should not be used as a guideline for how high the sintering temperature needs to be for the synthesis of the predicted UHTC compounds. Rather, $T^{*}$ should only be used as a guide quantity that indicates the anticipated synthesizability of the different compounds relative to each other. A more quantitatively sound equivalent to $T^{*}$ could be obtained from the free energy in the quasiharmonic approximation, but this would also suffer from the aforementioned issue with the electronic contribution, and it would be a much more computationally demanding quantity to obtain than $T^{*}$. 

\section{Conclusions}
We have demonstrated that the elastic properties of ultra-high temperature ceramic (UHTC) rocksalt carbides are highly tunable by composition. Deviations from the rule-of-mixtures approximation correlate strongly with lattice distortion, independent of whether the composition is low- or high-entropy. Further tuning is possible by moving from equimolar to non-equimolar compositions, particularly in high-entropy compounds with more compositional degrees of freedom. These insights were enabled by a fine-tuned MACE model (MACE-UHTC) that exhibits excellent precision in reproducing density functional theory reference data. Using this model, we identified a three-component mixture, HfCVCZrC, with a strong balance between synthesizability and a reduction in Young’s modulus. Screening the non-equimolar composition space of this mixture, we identified candidate compounds showing reductions up to 40 GPa relative to ZrC. These findings may result in substantial improvements to toughness in ultra-high temperature ceramic coating applications where existing single-metal UHTCs face prohibitive limitations.

\section{Methods}

\subsection{Data generation}
DFT training data, comprising total energies, forces, and stresses was generated using the Quantum ESPRESSO \cite{Giannozzi2009, Giannozzi2017, Giannozzi2020} package. The PBE\cite{PBE} exchange-correlation functional was used.
Pseudopotentials were chosen according to the validated recommendations of the `precision' flavor of the SSSP pseudopotential library\cite{Lejaeghere2016,Prandini2018}. Specifically, pseudopotentials of C and Nb were taken from versions 0.3.1\cite{PSLib031} and 1.0.0\cite{DalCorso2014} of the PSLibrary, respectively, while Hf was sourced from the Pseudo Dojo set\cite{vanSetten2018}, Ta and Zr were taken from version 1.2, and Ti and V from version 1.4 of the GBRV pseudopotential library\cite{Garrity2014}, respectively. The kinetic energy and charge density cutoffs were set to 50~Ry and 500~Ry, respectively, and we used a 24x24x24 k-point grid for primitive cell systems with appropriate scaling for supercells. These settings were sufficient to converge the bulk and shear moduli of the single-component UHTCs within 1~GPa. Marzari-Vanderbilt\cite{PhysRevLett.82.3296} smearing of electronic states was applied with a width of 20~mRy.

The training set consists of single-component UHTCs and mixture compounds in up to $4\times4\times4$ supercells of the FCC primitive cell. 
Cells were randomly strained with strain sampled from a normal distribution centered on 0\% with standard deviation 2\%. 
Similarly, atoms within the cells were rattled with Gaussian noise along each of the Cartesian axes with a mean overall displacement magnitude of 0.05~\AA. 
The Atomistic Simulation Environment (ASE) library \cite{HjorthLarsen2017} was utilized for constructing the supercells and generating the DFT data. 
The full dataset was then randomly divided via an 80\%-10\%-10\% train-valid-test split resulting in  4914, 614, and 613 configurations for training, validation, and test sets, respectively.

\subsection{Training the MACE-UHTC machine learning interatomic potential}

Training was carried out starting from the model weights of the general purpose MACE-MPA0 foundation model \cite{batatia2023foundation}, with all weights then permitted to vary during the training.
MACE-MPA0 has 128 invariant and 128 equivariant feature channels in two neural network layers with symmetry order $L=1$ and correlation order $\nu=3$. 
The atomic environment is sampled with a cutoff radius of 6~\AA.
For the finetuning, the batch size was set to 10 and the model was finetuned for a total of 400 epochs.
The ratio of energy, force, and stress weights was 1:100:1 for the first 200 epochs, then Stochastic Weight Averaging\cite{Izmailov_SWA_2018} was activated and the weights changed to 1000:100:1000 from then on, to improve the final values of the energy and, in particular, stress predictions.
Minimization of the loss function made use of the AMSGrad variant of the Adam\cite{Kingma2014} gradient optimizer, with a learning rate of 0.01 in the first stage of training, reduced to 0.0001 after 200 epochs.
Final root mean square (RMS) errors on the test set were 0.6~meV/atom, 19.6~meV\AA$^{-1}$, and 0.6~meV\AA$^{-3}$, for energies, forces, and stresses, respectively.

\subsection{Physical property prediction}

Crystal structures of UHTCs are optimized using our finetuned MACE-UHTC model in ASE\cite{HjorthLarsen2017} with a force tolerance of 0.001 meV/\AA, and using the Frechet cell filter to enable simultaneous optimization of the cell vectors in each configuration. The UHTC crystal structure is characterized by the mass density, the effective lattice parameter which is the average of the lengths of the three primitive cell vectors, and the displacement of the atoms in the lattice as compared to where they would sit in the rule of mixtures approximation.

We introduce a new, simplified descriptor of synthesizability, the effective stabilization temperature $T^{*}$, defined as the temperature where the mixing entropy is equal to the energy penalty arising from mixing the components together,

\begin{equation}
T^{*}=E_{\rm mix}^{\rm atom}/S_{\rm conf}^{\rm atom},
\end{equation}

\noindent where $S_{\rm conf}^{\rm atom}$ is the mixing entropy per atom,

\begin{equation}
S_{\rm conf}^{\rm atom}=-k_B \frac{N_{\rm metal}}{N_{\rm tot}} \sum_i x_i \ln x_i,
\end{equation}
\noindent where $\{x_i\}$ are metal fractions on a metallic sublattice, $N_{\rm metal}$ and $N_{\rm tot}$ are number of metals and the total number of atoms per cell, respectively. $S_{\rm conf}^{\rm atom}$ is computed from the mixing entropy per metallic site ($S_{\rm conf}^{\rm site}=-k_B \sum_i x_i \ln x_i$ ) by normalizing it per atom.

The mixing energy per atom, $E_{\rm mix}^{\rm atom}$, is computed as

\begin{equation}
E_{\rm mix}^{\rm atom}=E_{\rm conf}^{\rm atom}-\sum_i x_i E_{i},
\end{equation}

\noindent where $x_i$ is the concentration of component $i$ in the metallic sublattice, $E_i$ are the reference energies of binary carbides computed on an optimized primitive cell (containing one metal and one C atom), and $E_{\rm conf}^{\rm atom}$ is the total energy per atom in the mixture compound. Both energies are evaluated using the fine-tuned MACE-UHTC model. Although the configurational entropy $S_{\rm conf}^{\rm atom}$ does not change for ordered and disordered structures at the same composition (under the ideal-mixing assumption), the effect of ordering can still be observed through changes in the mixing energy $E_{\rm conf}^{\rm atom}$. Since our approach does not include vibrational and electronic contributions to the free energy, it should be emphasized that $T^{*}$ is not a quantitative prediction of the temperature required to synthesize the mixture compounds by for example sintering. It is an "effective" temperature that qualitatively indicates how difficult it is expected to be to create the mixture. The larger the $T^{*}$, the less likely that the compound can be synthesized as a single-phase crystal. $T^{*}$ is a descriptor that can be easily computed from the total energies predicted by MACE-UHTC for the optimized configurations considered, as such it takes account of the local distortions in the mixture compound and can be computed for any composition at a very affordable computational expense. It is also shown to correlate well to established, computationally much more expensive descriptors of synthesizability such as the DEED descriptor \cite{Divilov2024} in the SI.

To compute the elastic constants and other related mechanical properties of UHTCs, we construct the elastic tensor from stress data on the strained lattice\cite{M_de_Jong_2015}. We apply strain ($\pm 0.5$\% to stay within the elastic regime) individually in each independent component of the deformation matrix and compute the resulting stresses from which we construct the $6\times6$ elastic tensor $C_{ij}$. As the rocksalt crystal exhibits cubic symmetry, the only non-zero components are the $C_{11}$, $C_{12}$, and $C_{44}$ components, from which the bulk, shear, and Young's moduli are trivially calculated using the well-established Voigt-Reuss-Hill approach \cite{VRH_1967}.

In order to ensure accurate representation of the configurational disorder in the mixture compounds, ensemble averages of the mixtures are taken in a $6\times6\times6$ supercell which contains 432 atoms. 20 random configurations are generated for each equimolar compound, and 10 configurations for each non-equimolar mixture of HfCVCZrC, over which ensemble averages of converged relaxed structures are taken. The linear scaling of the computational cost of running trained MACE models with respect to the number of atoms in the crystal, and their excellent performance on GPU architectures, enable the computation of ensemble averages on such massive supercells, which would not be feasible with DFT. Such a large supercell not only ensures accurate representation of the random mixture, but enables use of the same supercell size for modelling both equimolar and non-equimolar mixtures.

In addition, binary mixtures, which are feasible to compute at DFT level in a small $2\times2\times2$ supercell with 10 configurations for each compound, have been investigated both with the MACE model and with DFT to enable a direct comparison of the MACE predictions with DFT for the elastic constants; this is presented in the SI.

\begin{acknowledgments}
This work was supported by the Hartree National Centre for Digital Innovation, a collaboration between STFC and IBM.
\end{acknowledgments}

\section*{Data availability}
The DFT training dataset and finetuned MACE-UHTC model generated during the current study, and the results calculated using the model and presented here, are available in the STFC eData repository at https://doi.org/10.5286/edata/960

 \newpage
 \onecolumngrid
 
\begin{center}
\textbf{\large Supplementary Information: Computational tuning of the elastic properties of low- and high-entropy ultra-high temperature ceramics}
\end{center}

\setcounter{equation}{0}
\setcounter{figure}{0}
\setcounter{table}{0}
\setcounter{section}{0}
\makeatletter
\renewcommand{\theequation}{S\arabic{equation}}
\renewcommand{\thesection}{S\arabic{section}}
\renewcommand{\thefigure}{S\arabic{figure}}
\renewcommand{\bibnumfmt}[1]{[#1]}
\renewcommand{\citenumfont}[1]{#1}

 \twocolumngrid
\section{Effective stabilization temperature - comparison to the DEED descriptor}

\begin{figure}
    \centering
    \includegraphics[width=\linewidth]{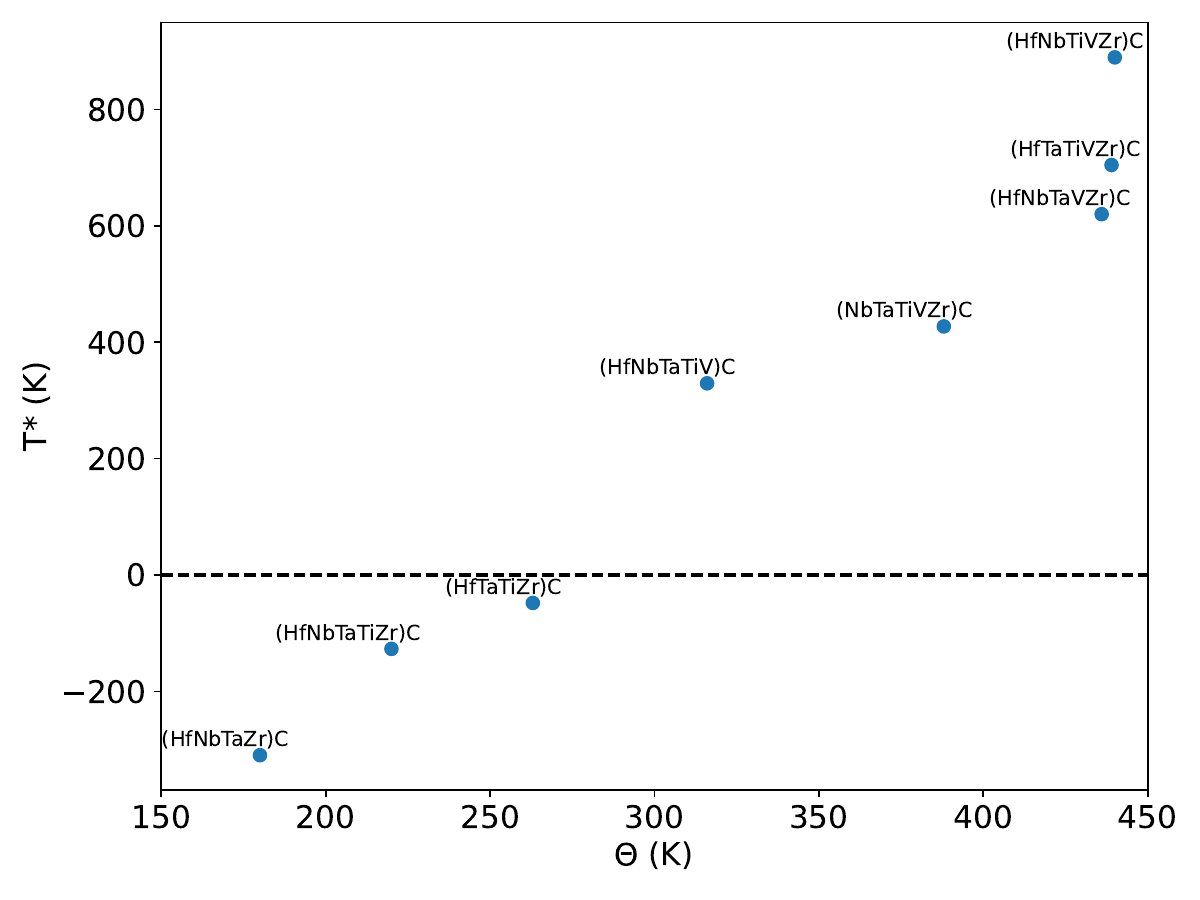}
    \caption{Comparison of the $T^{*}$ effective stabilization temperature with the DEED compensation temperature\cite{Divilov2024} for equimolar mixture compounds where both quantities are known.}
    \label{fig:Teff_comparison}
\end{figure}

The disordered enthalpy-entropy descriptor (DEED) was created in an effort to predict the synthesizability of high-entropy ceramics \cite{Divilov2024}. It is a quite expensive quantity to calculate that requires the thermodynamic density of states spectrum. Multiplied by Boltzmann's constant, its inverse takes on a temperature dimension, which is called the compensation temperature $\Theta$,

\begin{equation}
    \Theta = [k_B ({\rm DEED})]^{-1}.
\end{equation}

In Fig. \ref{fig:Teff_comparison} we show a plot of the effective stabilization temperature $T^{*}$ defined in the main manuscript, compared to the $\Theta$ values taken from literature \cite{Divilov2024}. We find reasonable correlation between the two quantities, which is particularly valuable as computing $T^{*}$ is much cheaper, especially when using an MLIP like MACE-UHTC.

\section{Comparison of elastic properties predicted by finetuned MACE with DFT in small supercells}

\begin{figure}
    \centering
    \includegraphics[width=\linewidth]{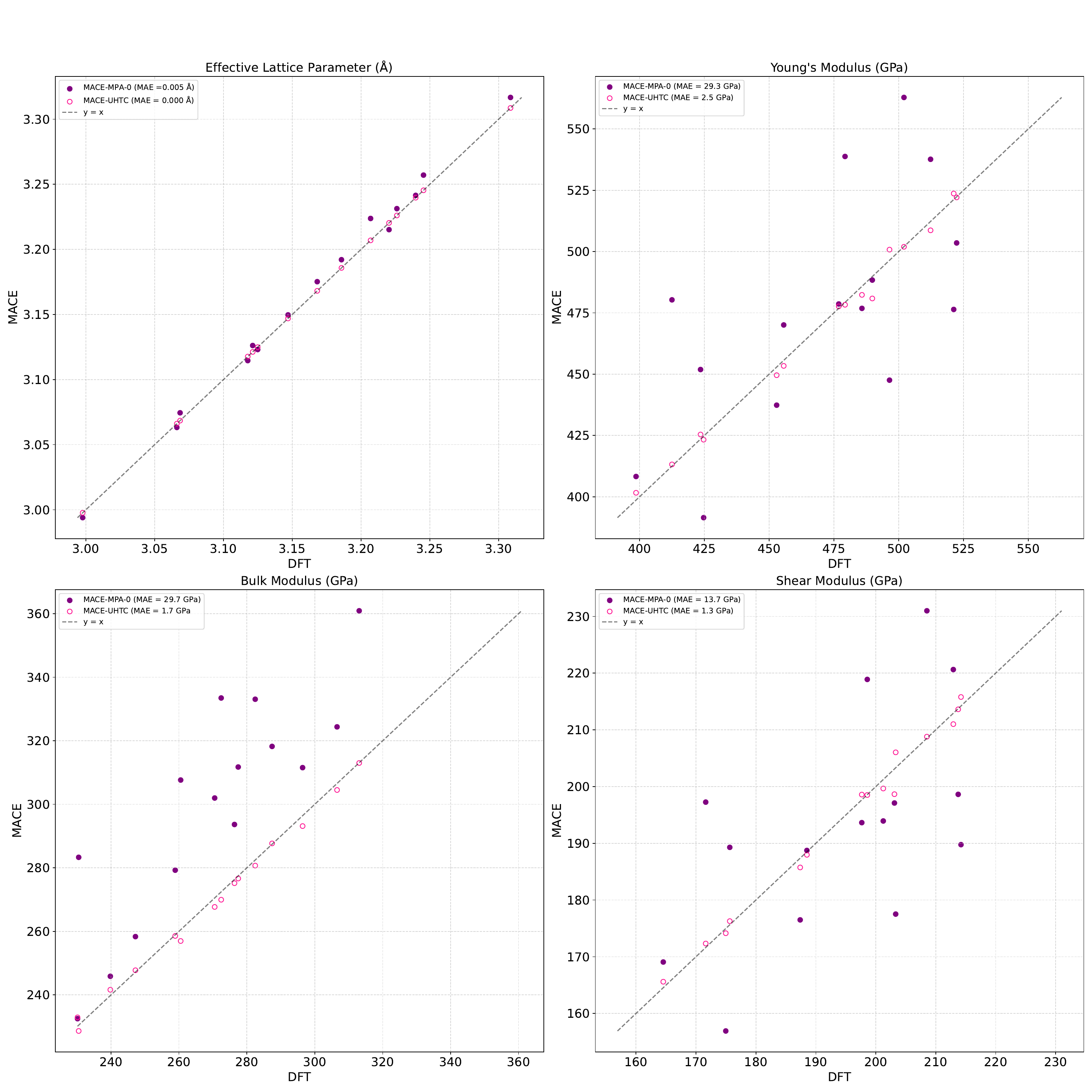}
    \caption{Predictions by the MACE-MPA-0 model and our fine-tuned MACE-UHTC model for binary mixtures of UHTCs compared to DFT reference for the lattice parameter, and the Young's, shear, and bulk moduli.}
    \label{fig:models_vs_DFT}
\end{figure}

Fig. \ref{fig:models_vs_DFT} shows how well our finetuned MACE-UHTC model compares to DFT reference data of the lattice parameter and the elastic moduli for 2-component UHTCs in $2\times2\times2$ supercells. Data points cluster very close to the diagonal, aligned with the performance found during training of the model.

For comparison, we also show the predictions for these compounds produced by MACE-MPA-0 prior to finetuning. The lattice parameters are reasonably reproduced, however, all of the moduli exhibit substantial errors, in particular the bulk modulus which is consistently overestimated. This clearly demonstrates how essential it is to finetune a MACE model before attempting to make predictions for measurable physical properties such as elastic constants.

\section{Overview of physical properties}

\begin{figure*}
    \centering
    \includegraphics[width=0.9\linewidth]{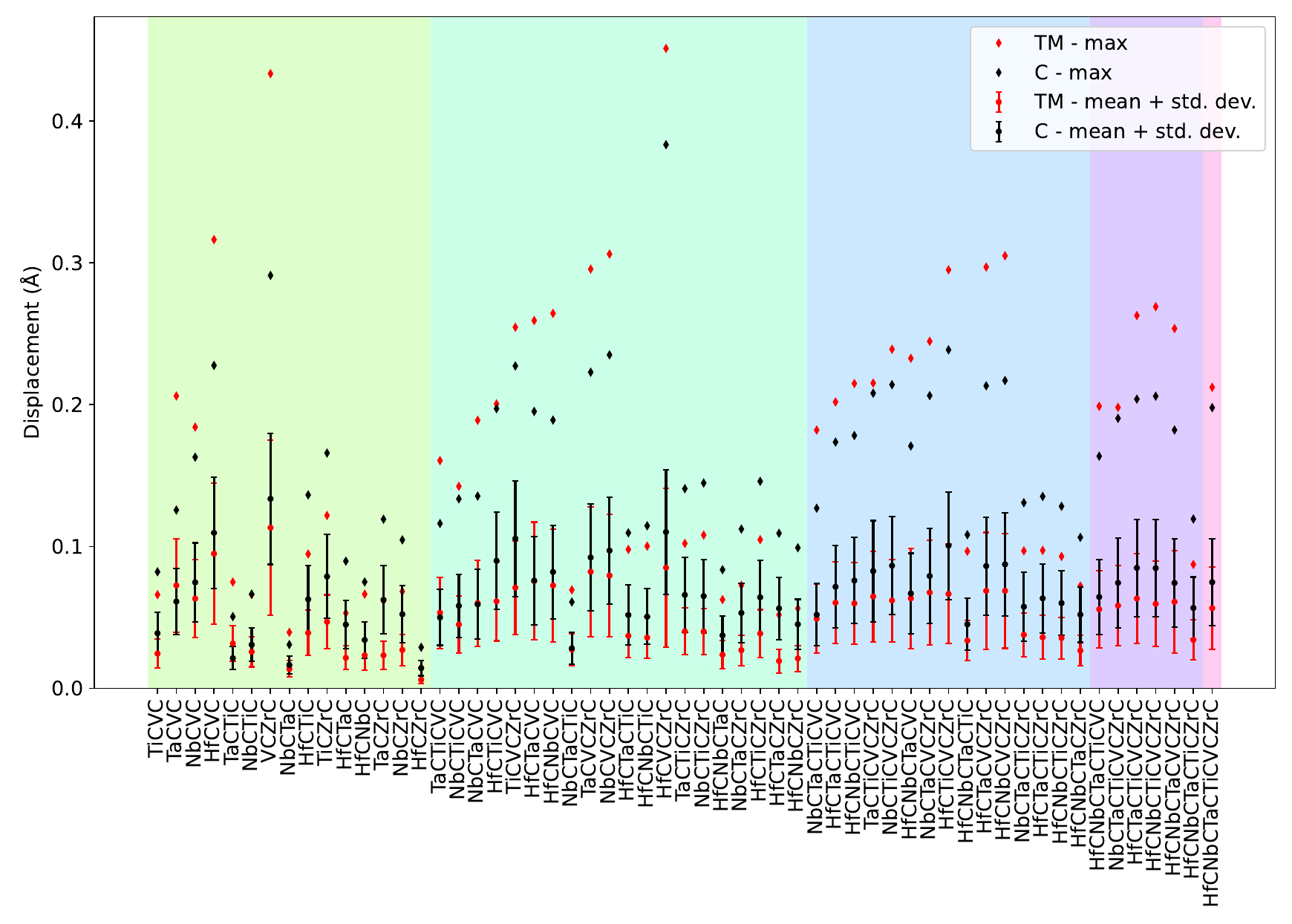}
    \caption{Average, spread (standard deviation), and maximum value of the ionic displacement on the carbon (C) and the transition metal (TM) sublattice averaged over the generated ensemble of random configurations in each mixture compound.}
    \label{fig:position_distortion}
\end{figure*}

Fig. \ref{fig:position_distortion} shows how much the atoms displace in each equimolar mixture compound, considering all random configurations. Despite the C atoms being considerably lighter, their sublattice only exhibits slightly larger displacement than the transition metal sublattice. Moreover, the maximum value of displacements within the ensemble is larger on the metal sublattice.

\begin{figure*}
    \centering
    \includegraphics[width=0.9\linewidth]{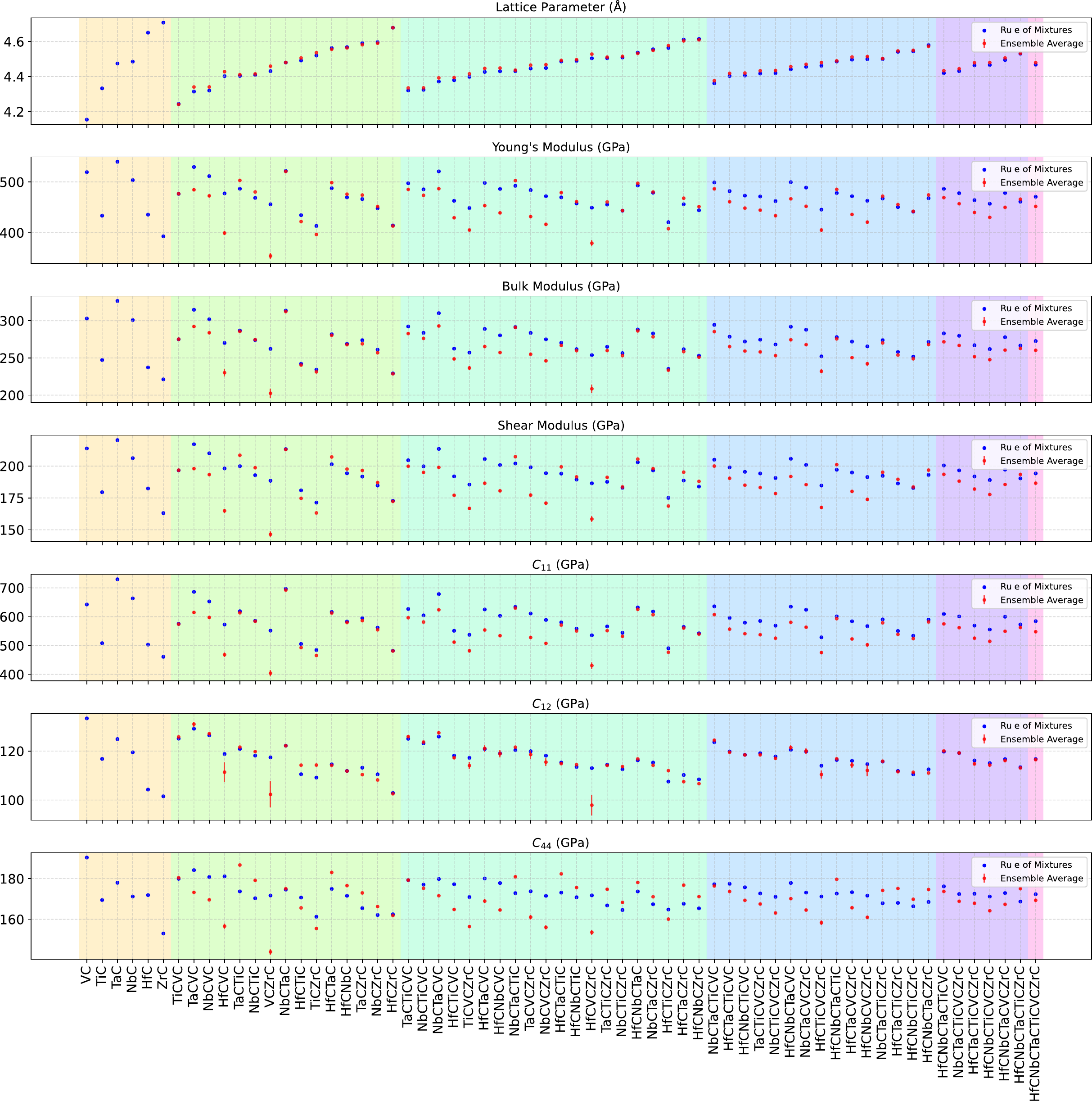}
    \caption{Overview of physical properties of equimolar UHTC compounds, showing the standard deviation of the ensemble averaging as error bars around the mean ensemble averaged values.}
    \label{fig:666_moduli_sort_latt}
\end{figure*}
The distortion of the cell on the other hand is much less pronounced. 
We analyzed this by computing the strain matrices which transform the ideal cubic lattice vectors into the optimized lattice vectors. We found that the off-diagonal strain is negligible, and that the diagonal components are largely isotropic, which is to say, the cubic symmetry is effectively retained; this is further confirmed by analyzing the full elastic tensor, where all components which should be exactly zero in a cubic crystal, are less than 1 GPa, over two orders of magnitude below $C_{11}$, $C_{12}$, and $C_{44}$. Therefore, the simplest way to quantify the lattice distortion is in fact through the lattice parameter, which is shown in Fig. \ref{fig:666_moduli_sort_latt} comparing the ensemble average to the rule of mixtures approximation. The data points are grouped by number of components, and each group is sorted by increasing mean lattice parameter of the component UHTCs. Fig. \ref{fig:666_moduli_sort_latt} also shows the finetuned MACE-UHTC model predictions of the elastic constants and moduli for every mixture compound, comparing ensemble averages to the rule of mixtures approximation.
\clearpage
\bibliographystyle{apsrev4-1}
\bibliography{references}

\end{document}